\documentstyle[e-e-ijmpa,twoside,psfig]{article}

\renewcommand{\thefootnote}{\fnsymbol{footnote}}        

\begin{document}

\normalsize\textlineskip
\pagestyle{empty}

\title{PHYSICS AT \boldmath$e^{-} e^{-}$: A CASE FOR MULTI-CHANNEL STUDIES}

\author{HITOSHI MURAYAMA%
\footnote{This work was supported in part by the U.S.  Department of
  Energy under Contracts DE-AC03-76SF00098, in part by the National
  Science Foundation under grant PHY-95-14797, and also by Alfred P.
  Sloan Foundation.}}

\address{Theoretical Physics Group\\
     Ernest Orlando Lawrence Berkeley National Laboratory\\
     University of California, Berkeley, California 94720}
         
\address{Department of Physics\\
     University of California, Berkeley, California 94720} 

\maketitle

\begin{abstract}
  I argue that it would be crucial to have as many channels as
  possible to understand the physics of electroweak symmetry breaking
  (EWSB) in next-generation collider experiments.  A historic example
  of the parity violation and the $V-A$ interaction is used to make
  this point.  An $e^- e^-$ option offers us a new channel in this
  respect.  The usefulness of this channel is exemplified for the case
  of supersymmetry and of the strongly coupled EWSB sector.
\end{abstract}
\pagestyle{empty}
\setcounter{footnote}{0}
\renewcommand{\thefootnote}{\alph{footnote}}

\vspace*{1pt}\textlineskip      
\section{Why are we here?}

So, here is another workshop on collider physics.  Specifically on a 
rather exotic collider option, $e^{-} e^{-}$ collider.  Why are we 
doing this, after all?

The answer to this question is quite simple.  We believe that the 
physics of electroweak symmetry breaking (EWSB) is the most pressing 
question in particle physics.  And it is going to be a challenging 
task to completely reveal all secrets of EWSB. It will take 
substantial experimental and theoretical efforts to understand it.  
For this aim, having as many possible channels as possible will  
probably be necessary.

Why multi-channel?  We heard about the complementarity of hadron and 
lepton machines so many times.  Maybe enough of it.  And we are 
talking here about yet another possible collider option.  Why bother?

I would like to remind you of an example in the history of particle
physics where it was crucial to attack the same problem from many
different channels.  It is the $V-A$ form of the charged-current weak
interaction.\footnote{Of course, all of this happened before I was
  born; I rely completely on reviews,\cite{review} a random reading
  of conference proceedings and private communications.  There must
  have been much more behind-the-scene confusions than I briefly
  outline here.}

The first hint for parity violation came from a purely hadronic
process.  In cosmic ray and beam-based studies of strange particles,
there appeared two particles with opposite parities but with exactly
the same mass and lifetime---the famous $\tau$-$\theta$ puzzle.
And it took geniuses like T. D. Lee and C. N. Yang (1956) to make a bold
step towards the resolution: the parity violation.  They pointed out
that even though there was a wealth of evidence that  parity was a
good symmetry of the electromagnetic and strong interactions, there was
basically no experimental test whether the weak interaction preserves parity.
If it does not, we can identify the $\tau$ and  $\theta$ mesons which
decay into states with  opposite parities.

The evidence for parity violation came from two different channels in
1957: semi-leptonic and purely leptonic.  The semi-leptonic channel is
the experiment on $^{60}$Co $\beta$ decay (C. S. Wu {\it et al}\/).
The correlation between directions of the applied magnetic field and
the  electron momentum established the violation of parity.
The purely leptonic channel is the experiment by Garwin, Lederman and
Weinrich, using the sequence of decays $\pi \rightarrow \mu \rightarrow
e$.  The angular distribution of the $\mu \rightarrow e$ decay showed
the stopped muon was highly polarized.  It is interesting that both
papers were published in the same volume of {\sl Phys.\ Rev.}\/ {\bf
  105} side by side.

The violation of parity opened up a big confusion in the community.
The Fermi Hamiltonian of the weak interaction had to be reexamined by
allowing all possible 40 independent parameters of scalar, vector and
tensor Lorentz structures.  It also appeared, before 1957, that the
scalar and tensor interactions are dominant.

In order to choose $V$, $A$ over $S$, $T$ and $P$, different types of
nuclei had to be used.  In the study of the $(\beta - \nu)$ angular
correlation, different nuclei sit at different points on the so-called
Scott diagram, depending on the relative magnitude of Fermi and
Gamov--Teller transitions and recoil energy spectra.  The data using
$^{19}$Ne can be explained either by a combination of $V, A$ or of $S,T$.
However, the data using the $^{35}$Ar nucleus prefer $V$.  The data from
$^{23}$Ne and $^{6}$He prefer $A$.  Only after putting all of them
together, the choice of $V$ and $A$ comes out preferred over $S$ and
$T$.  This analysis could not be done without the determination of the
negative helicity of the neutrino, evidence for which came from the $K$
capture of $e^- \hbox{Eu}^{152m} \rightarrow \nu \hbox{Sm}^{152*}$ by
Goldhaber, Grodzins, and Sunyar in 1958, which reduced the number of
parameters by a factor of two.  However, none of these measurements
was able to establish precisely the $V-A$ form, because the strong
interaction renormalizes the axial-vector coupling even though the CVC
hypothesis keeps the normalization of the $V$ part non-%
renormalized.\footnote{The PCAC, however, related the axial-vector
  coupling to the pion-nucleon coupling (Goldberger--Treiman relation)
  in 1958.}

In the purely leptonic channel,  muon decay was studied in detail.
By fitting the energy spectrum and angular correlation with polarized
muons, the parameter space was restricted.  The Michel parameter
$\rho$ and the asymmetry parameter $\xi$ chose the $V-A$ theory by
1960.  Even at the present time, this analysis still offers the best
evidence for the $V-A$ nature.  The 20 parameters were reduced to just
one.

And throughout all this, the universality of the weak interaction 
between muon decay (purely leptonic) and $\beta$-decay
(semi-leptonic) played its role as the backbone of the development.

I am telling students in my particle physics course that the
discovery of the $V-A$ nature led to a major paradigm shift in the way
we understand the elementary particles.  This is not much emphasized,
but it is a fundamental change.  Since the days of Pauli, the spin of the
electron used to be an {\it additional}\/ degree of freedom attached
to a non-relativistic electron.  However, the $V-A$ nature forced us
to go to a totally different idea.  The distinction of the left-handed
and right-handed helicity states is fundamental; these are {\it
  different}\/ particles!  And this fundamental distinction does not
allow the electron to stop, because then the helicity would lose its
meaning.  The elementary particles are {\it chiral}\/.  And only
because of the Higgs boson condensate, the left-handed and
right-handed states can convert to each other and the electron can
peacefully ``sit'' at rest.

I anticipate that the experimental study of the physics of EWSB will 
be as confusing and complex.  In order to fully reveal all the 
secrets, we are likely to need as many channels as possible.  This is 
why we would like to have an $e^{+} e^{-}$ machine in addition to the 
already-approved LHC project.  As an added bonus, an $e^{+} e^{-}$ 
machine will offer an $e^{-} e^{-}$ option at  minimum cost and
give us a handle on yet another channel in the study.

Of course, at this point we do not know how having different channels
will help us to understand the physics of EWSB, because we do not yet
know what the physics is.  But we can do a case study using the
known proposed mechanisms of EWSB and associated new particles 
in order to study how
an $e^{-} e^{-}$ option would help us.  In the next 
sections we will see the cases of supersymmetry (weakly interacting 
EWSB) and strongly interacting EWSB from this point of view.

\vspace{1pc}
\section{Characteristics of \boldmath$e^{-}e^{-}$ Experiments}

In a review on $e^+ e^-$ physics,\cite{annual} Michael Peskin and I
listed the three important  
characteristics of an $e^{+} e^{-}$ collider: cleanness, ``holism,'' 
and democracy.  The cleanness refers to the fact that the event rate 
at an $e^{+} e^{-}$ machine is low and offers a desirable environment 
for detectors and physics.  Since the initial state is well-defined, 
including the polarization of at least the $e^-$ beam, we know what 
we are doing.
And the background is calculable.  The ``holism'' refers to the 
capability to capture the entire event.  Since the kinematics is 
well-defined, we can use the beam energy constraint to facilitate the
reconstruction of the event.   Lastly, democracy means that all
final states, both signal and background, have comparable cross
sections;  hence, the experiment is suitable for studying many
different types of particles and interactions simultaneously.

By switching to the $e^{-} e^{-}$ option, we retain cleanness and
``holism'', but we lose democracy.  Basically, almost all final states
{\it cannot}\/ be produced, both the signals and the backgrounds.
There is no annihilation between $e^{-}$ and $e^{-}$,\footnote{A
  possible di-lepton resonance is an exception.}\,  which makes the
level of the background even lower.  However, this opens up a new
channel in the experiment.  There are some signals which can be
studied with this option---there may be some exotic and/or rare final
states.  I will give a few relevant examples later.  As emphasized
earlier, we do not know what specifically would be the best signal to 
study in this new channel.  But it is likely that there are some signals
arising from the complexity in the physics of EWSB.

\vspace{1pc}
\section{Supersymmetry}

Let us take the superpartner of the electron, the selectron, for the
purpose of this discussion.  At the LHC, the selectron can be
studied in detail if it appears in the decay of the second 
neutralino, $\tilde{\chi}_{2}^{0} \approx \tilde{W}^{0}$, with its
possible decay into $\tilde{e}^{\pm} e^{\mp}$.  It is not always 
possible to study the selectron, however, depending on the precise 
pattern of the superparticle spectrum.

At an $e^{+} e^{-}$ machine, the study of the selectron is 
more-or-less trivial as long as it exists within kinematic reach.
Furthermore, the use of a polarized right-handed electron beam
suppresses the $W$-pair background, and many details, such as the mass
of the selectron, the neutralino in its decay product, and the 
electron-selectron-neutralino coupling can be studied.

\begin{figure}
\centerline{\psfig{file=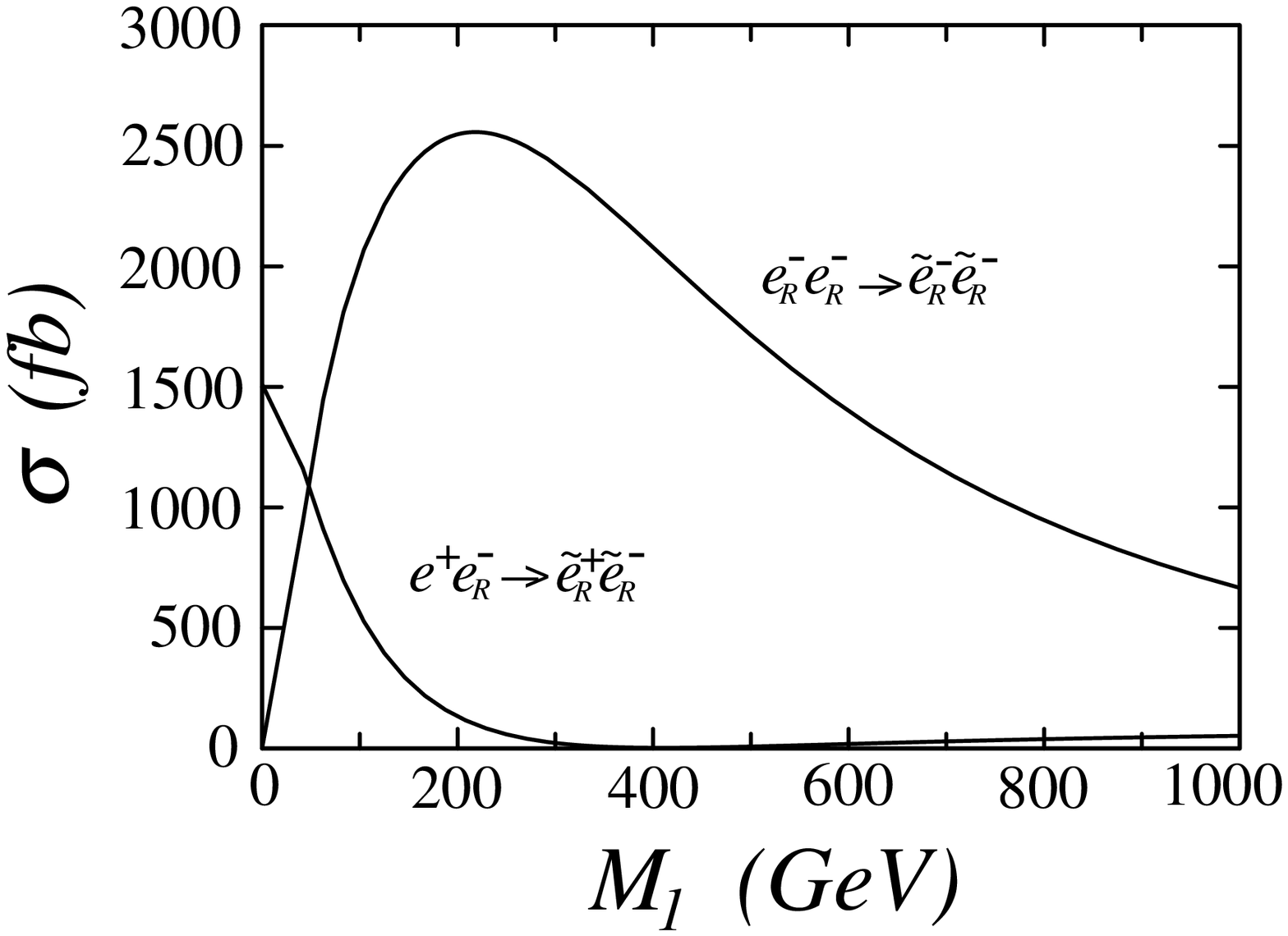,width=0.6\textwidth}}
\fcaption{The total selectron pair production cross
  sections for the $e_R^- e_R^-$ and $e^+ e_R^-$ modes with
  $m_{\tilde{e}_R}=150$~GeV and $\protect\sqrt{s}=500$~GeV, as
  functions of the Bino mass $M_1$.\cite{CFP}\label{M1dependence}}
\end{figure}

At an $e^{-} e^{-}$ collider, the selectron can be studied in even
greater detail.  One reason is that the destructive interference
between $s$-channel $\gamma$, $Z$ exchange and $t$-channel
$\tilde{B}$ exchange in the $e^{+} e^{-}$ annihilation is gone in
$e^{-} e^{-}$ collisions, because of the absence of the $s$-channel
diagram.  This results in larger cross sections (see
Fig.~\ref{M1dependence}).  Second, there is no 
$W$-pair background even with the left-handed electron beam.  Third, 
one can control the polarization of both of the beams, which can
effectively turn on or off the final states of interest.  The $e^{-}_{L}
e^{-}_{R}$ initial state produces only the $\tilde{e}^{-}_{L} 
\tilde{e}^{-}_{R}$ final state.  Similarly, $e^{-}_{L} e^{-}_{L}$ leads 
to $\tilde{e}^{-}_{L} \tilde{e}^{-}_{L}$ and $e^{-}_{R} e^{-}_{R}$ to 
$\tilde{e}^{-}_{R} \tilde{e}^{-}_{R}$.  The experimental verification 
of this simple selection rule would tell us that the scalar particles 
can carry chirality, which is the very reason why supersymmetry 
protects scalar masses against radiative corrections.  And the 
threshold behavior of $e^{-}_{R} e^{-}_{R} \rightarrow 
\tilde{e}^{-}_{R} \tilde{e}^{-}_{R}$ is $\propto~\beta$, as opposed to
the $\beta^{3}$ behavior in $e^{+} e^{-}$ annihilation. This is suitable
for an accurate determination of the selectron mass.

\begin{figure}
        \centerline{\psfig{file=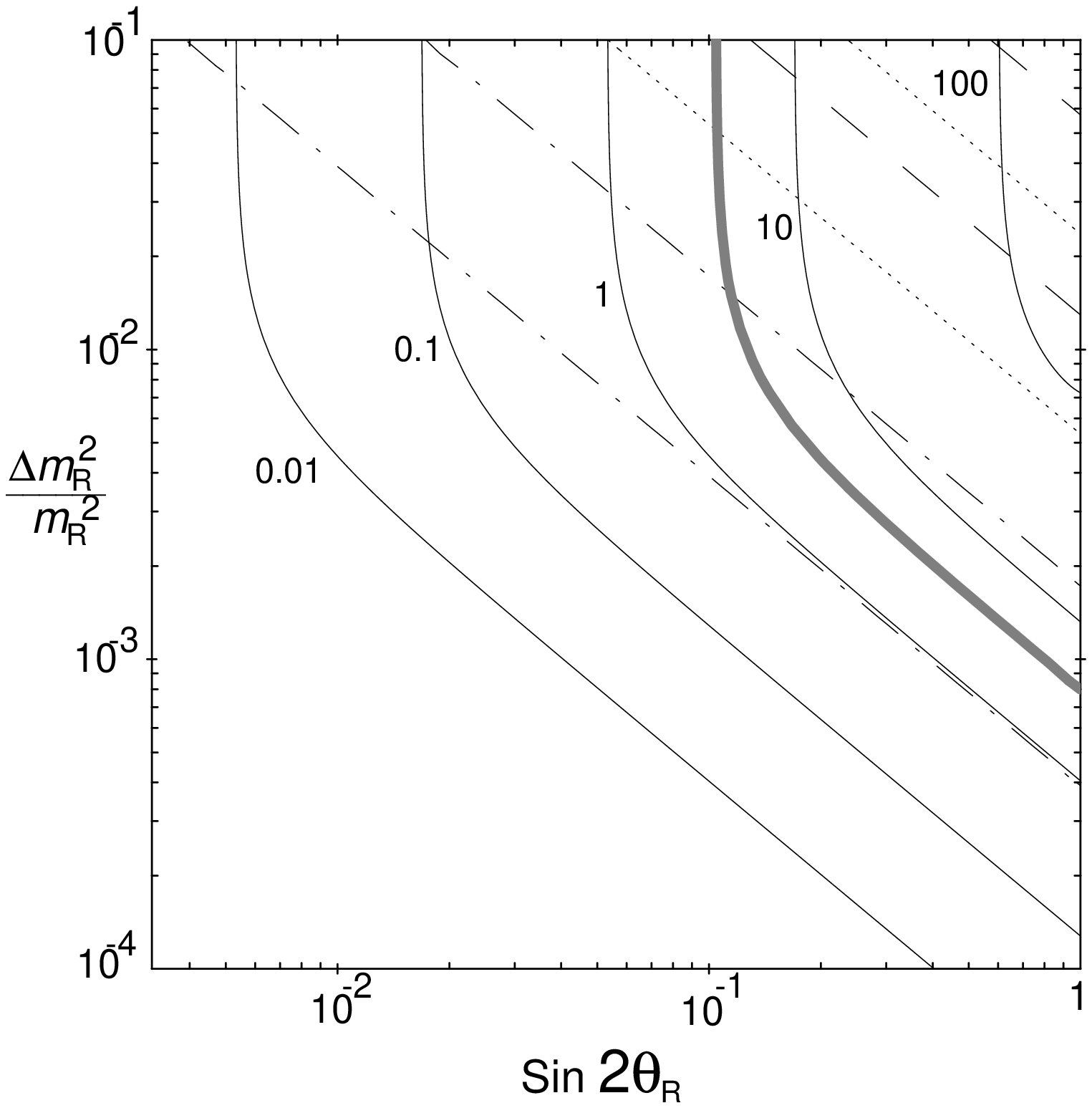,width=0.45\textwidth}
         \hfill
        \psfig{file=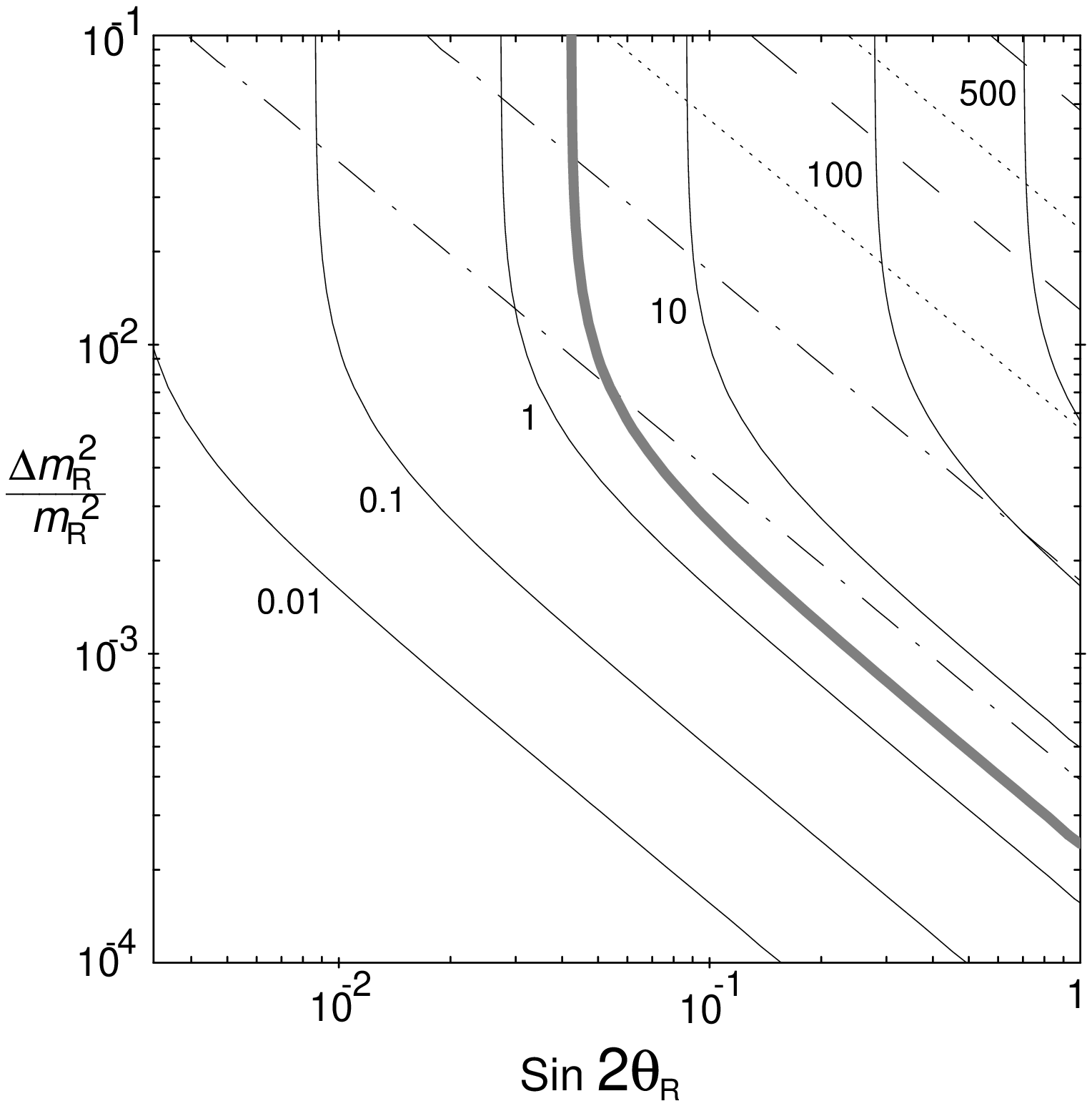,width=0.45\textwidth}}
        \fcaption{Contours of constant $\sigma (e^+e^-_R\to
          e^{\pm} \mu^{\mp} \tilde{\chi}^0\tilde{\chi}^0 )$ (left) and
          $\sigma (e^-_R e^-_R \to e^- \mu^-
          \tilde{\chi}^0\tilde{\chi}^0 )$ (right) in fb for the NLC, with
          $\protect\sqrt{s} = 500~\hbox{GeV}$, $m_{\tilde{e}_R},
          m_{\tilde{\mu}_R} \approx 200~\hbox{GeV}$, and $M_1 =
          100~\hbox{GeV}$ (solid).  The thick gray contour represents
          the experimental reach in one year.  Constant contours of
          $B(\mu \to e\gamma)$ are also plotted for left-handed
          sleptons degenerate at 350 GeV.\cite{ACFH}\label{lfvfig2}}
\end{figure}

The absence of the $W$-pair background and the increase in the cross
section allows us to study rare processes better than in an $e^{+}
e^{-}$ collision.  One interesting example will be discussed by
Jonathan Feng later in this workshop.  What he pointed out together
with Nima Arkani-Hamed, Hsin-Chia Cheng, and Lawrence Hall, is that the
selectron may have a small mixing with the smuon; this mixing then
results in a
phenomenon similar to the neutrino oscillation.  When, for instance,
$\tilde{e}^{+} \tilde{e}^{-}$ is produced from $e^{+} e^{-}$
annihilation, the produced $\tilde{e}$ is in its ``interaction
eigenstate'', which may differ from its mass eigenstate.  Then the
$\tilde{e}$ can oscillate to a mixture of $\tilde{e}$ and
$\tilde{\mu}$ and can decay into muon as well.  This results in a
final state of $e^{\pm} \mu^{\mp} \not\!\!E$.\cite{ACFH} A search for
this phenomenon can be done quite well in the $e^{+} e^{-}$
environment, but much more efficiently in  $e^{-} e^{-}$ collisions,
as seen in Fig.~\ref{lfvfig2}, because of the absence of the $W$-pair
background and of the higher cross section.

Another important example is the following precision measurement:  
By studying the
selectron in detail, the $\tilde{e}$-$e$-$\tilde{B}$ coupling can be
measured precisely in the $e^{+} e^{-} \rightarrow \tilde{e}^{+}
\tilde{e}^{-}$ process.\cite{NFT} Again, thanks to the lower background
and the higher cross section, this measurement can be done better with
an $e^{-} e^{-}$ option.\cite{CFP} Since this coupling is supposed to
be the same as the $U(1)_{Y}$ gauge coupling $g' = g/\cos\theta_{W}$
because of supersymmetry, this would be an important quantitative test
if the interactions preserve supersymmetry.  If a small violation
were seen, it could be interpreted as the violation of
supersymmetry due to the heaviness of squarks---which can modify the
$\tilde{e}$-$e$-$\tilde{B}$ coupling at the one-loop level.  The small
violation could be used to determine roughly how massive the squarks
are.  This effect is called a super-oblique correction, in analogy to
the determination of the top quark and Higgs boson masses from
electroweak precision measurements.  The better accuracy achievable at
an $e^{-} e^{-}$ option could prove essential in this type of study.

One may also look for some exotic physics with an $e^{-} e^{-}$ 
option.  For instance, the $R$-parity violating interactions
$$
   \lambda_{132} L_{1} L_{3} \mu^{c} + \lambda_{231} L_{2} L_{3} e^{c}
$$
do not generate $\mu \rightarrow e\gamma$ or $\mu$-$e$ conversions
because of the conserved $L_{e} + L_{\mu}$ quantum number.  The best
bound on these couplings come from muonium conversion, $\mu^{+}
e^{-} \leftrightarrow \mu^{-} e^{+}$,  and  $e$-$\mu$-$\tau$
universality; it is of the order of 0.1 for $m_{\tilde{L}} \sim 200$~GeV.
These interactions can cause the reaction $e^{-}_{L} e^{-}_{R}
\rightarrow \mu^{-}_{L} \mu^{-}_{R}$ with essentially no background.
The event rate is given roughly by
$$
        \frac{\#(\hbox{event})}{20~\hbox{fb}^{-1}} \simeq
        10^{6} \times \lambda^{4} \times 
        \left(\frac{200~\hbox{GeV}}{m_{\tilde{L}}}\right)^{4}
        \left( \frac{\sqrt{s}}{200~\hbox{GeV}}\right)^{2} .
$$
One can see more than 5 events if $\lambda \geq 0.05$, which is
below the current limits.

\vspace{1pc}
\section{Strongly-Interacting EWSB Sector}

The strongly-interacting EWSB sector will pose a great challenge at
next-gen\-eration collider experiments.  The signal is a rather
featureless enhancement in the interaction between $W$- and $Z$-bosons 
at very high energies, $\sqrt{\hat{s}} \geq 1$~TeV. The TeV-scale 
experiments in this case are regarded as the ``low-energy limit'' of 
the true dynamics of EWSB, which is analogous to pion scattering in
the low-energy limit of the QCD, described by the chiral Lagrangian.  A
better manifestation of the dynamics of EWSB may show clearer at yet 
higher energies, such as at a 4~TeV muon collider.  Until we can reach 
such a high center-of-mass energy, all we can do is to study the 
``low-energy'' interaction of $W$-bosons in detail and speculate on 
the dynamics.  For this purpose, it is necessary to determine the 
size of the interaction (scattering lengths) in all possible 
channels.  Table~\ref{Barklow} shows the relative merit of
different channels for different scenarios.  

\begin{table}
        \tcaption{Statistical significances of strong EWSB signals at
        the NLC and LHC.\cite{Barklow-Iwate}\label{Barklow}}
\vskip3pt
\begin{center}
        \begin{tabular}{|l|l|c|c||c|c|c|}
                \hline
                Collider & Process & $\sqrt{s}$ & ${\cal L}$ & $M_{\rho} = $ &
                $M_{H} =$ & LET\\
                & & (TeV) & (fb$^{-1}$) & 1.5~TeV & 1~TeV & \\ \hline
                NLC & $e^{+}e^{-} \rightarrow W^{+} W^{-}$ & .5 & 80 & 
                7$\sigma$ & $-$ & $-$\\
                NLC & $e^{+} e^{-} \rightarrow W^{+} W^{-}$ & 1.0 & 200 & 
                35$\sigma$ & $-$ & $-$ \\
                NLC & $e^{+} e^{-} \rightarrow W^{+} W^{-}$ & 1.5 & 190 & 
                366$\sigma$ & $-$ & 5$\sigma$ \\
                NLC & $W^{+} W^{-} \rightarrow ZZ$ & 1.5 & 190 & $-$ & 22$\sigma$ & 
                8$\sigma$ \\
                NLC & $W^{-} W^{-} \rightarrow W^{-} W^{-}$ & 1.5 & 190 & $-$ & 
                4$\sigma$ & 6$\sigma$ \\
                \hline
                LHC & $W^{+} W^{-} \rightarrow W^{+} W^{-}$ & 14 & 100 & $-$ & 
                14$\sigma$ & $-$\\
                LHC & $W^{+} W^{+} \rightarrow W^{+} W^{+}$ & 14 & 100 & $-$ & 
                3$\sigma$ & 6$\sigma$ \\
                LHC & $W^{+} Z \rightarrow W^{+} Z$ & 14 & 100 & 
                7$\sigma$ & $-$ & $-$\\
                \hline
        \end{tabular}
\end{center}
\end{table}

One possibility which has not been studied is the $ZZ\rightarrow ZZ$
channel.  Due to a magical cancelation, there is no strong
interaction in this channel according to the Low-Energy Theorem (LET).
However, there is likely to be a strong interaction turning on at
order $(s/v^{2})^{2}$ and hence it uniquely picks up the
model-dependent piece at the next-to-leading order in the derivative
expansion. The feasibility of this study has to be examined.  It is
certain, however, that the $e^{-} e^{-}$ option is best suited for
this study, using right-handed electron beams $e^{-}_{R} e^{-}_{R}
\rightarrow e^{-}_{R} e^{-}_{R} ZZ$, because of the absence of the $WW$
fusion mechanism.

\vspace{1pc}
\section{Many More}

I will not go into the other possible interests in the $e^{-} e^{-}$ 
option as discussed in the literature, because they are covered by other
speakers in this workshop.  It includes the Higgs production from 
$ZZ$ fusion (Minkowski), $t$-channel $Z'$-exchange (Rizzo), 
doubly-charged Higgs $H^{--}$ from $W^{-} W^{-}$ fusion (Gunion), 
$H^{-} H^{-}$ production from $W^{-} W^{-}$ fusion (Haber), 
supersymmetry signatures (Peskin, Thomas, Feng, Cheng), strong EWSB 
(Han), anomalous triple-gauge-boson vertices (Choudhury), dilepton
resonance (Frampton), compositeness (Barklow), leptoquarks (Rizzo), 
$e^{-} e^{-} \rightarrow W^{-} W^{-}$ from right-handed Majorana 
neutrino $t$-channel exchange (Greub, Min\-kowski, Heusch),
$\gamma\gamma \rightarrow t\bar{t}$ (Hewett), $\gamma\gamma 
\rightarrow H$ (Takahashi).  Many of the signatures employ the new 
channel available only in an $e^{-} e^{-}$ collision.

\vspace{1pc}
\section{Conclusions}

The physics of EWSB is likely to produce rather messy and confusing
data.  In order to sort these out, we would like to have as many channels
as possible.  This is the most convincing argument behind the 
complementarity between the LHC and an $e^{+} e^{-}$ linear collider.  
A further extrapolation of this argument raises the interest in the
$e^{-} e^{-}$ option as a natural sibling to the $e^{+} e^{-}$ machine.
It offers new channels and hence new observables.

We do not know how exactly data from various channels will collude
to reveal the secrets of the physics of EWSB, because we do not know
the physics yet.  But in many examples that we know, the $e^{-} e^{-}$
option offers new, interesting, and valuable observables.

\vspace{1pc}
\nonumsection{References}


\begin{thebibliography}{99}
\bibitem{review} T. D. Lee and C. S. Wu, {\it Ann. Rev. Nucl. Sci.}
  {\bf 15}, 381 (1965); R. N. Cahn and G.~Goldhaber, ``The
  Experimental Foundation of Particle Physics,'' Cambridge University
  Press, 1989.
\bibitem{annual} H. Murayama and M. E. Peskin, {\it
  Ann. Rev. Nucl. Part. Sci}\/, {\bf 46}, 533 (1996).
\bibitem{CFP} H.-C. Cheng, J. L. Feng, and N.
  Polonsky, {\it Phys. Rev.}\/ {\bf D56}, 6875 (1997).
\bibitem{ACFH} N. Arkani-Hamed, H.-C. Cheng, J. L. Feng, and
  L. J. Hall, {\it Phys. Rev. Lett.}\/ {\bf 77}, 1937 (1996).
\bibitem{NFT} M. M. Nojiri, K. Fujii, T. Tsukamoto, {\it
  Phys. Rev.}\/ {\bf D54}, 6756 (1996).
\bibitem{Barklow-Iwate} T. L. Barklow, plenary talk presented at
  3rd International Workshop on Physics and Experiments with Linear
  Colliders, Morioka-Appi, Iwate, Japan, Sep 8--12, 1995,
  Proceedings ed. by A. Miyamoto, Y. Fujii, T. Matsui, and S. Iwata,
  World Scientific, Singapore, 1996.
\end{thebibliography}
\end{document}